\documentclass[aps,preprint,showpacs,preprintnumbers,amsmath,amssymb]{revtex4}

\usepackage{dcolumn}
\usepackage{mathrsfs}
\usepackage{bm}
\usepackage{amsmath,amssymb,epsfig,float}

\begin{document}

\title{Projective measurements and generation of
entangled Dirac particles in Schwarzschild Spacetime}

\author{Jieci Wang, Qiyuan Pan and Jiliang Jing\footnote{Corresponding author, Email: jljing@hunnu.edu.cn}}
\affiliation{ Institute of Physics and  Department of Physics,
\\ Hunan Normal University, Changsha, \\ Hunan 410081, P. R. China
\\ and
\\ Key Laboratory of Low-dimensional Quantum Structures
\\ and Quantum
Control of Ministry of Education, \\ Hunan Normal University,
Changsha, Hunan 410081, P. R. China}

\vspace*{0.2cm}
\begin{abstract}
\vspace*{0.2cm}  It is shown that the projective measurements made
by Bob who locates near the event horizon of the Schwarzschild black
hole will create entangled particles detected by Alice who stays
stationary at the asymptotically flat region. It is found that the
degree of entanglement decreases as the frequency of the detected
particles increases and approaches to zero as the frequency
$\omega_\mathbf{k} \rightarrow\infty$. It is also noted that the
degree of entanglement increases as the Hawking temperature
increases. Especially,  the particle state is unentangled when the
Hawking temperature is zero and approaches a maximally entangled
Bell state when the black hole evaporates completely.

\end{abstract}

\vspace*{1.5cm}
 \pacs{03.65.Ud, 03.67.Mn, 04.70.Dy,  97.60.Lf}

\maketitle

\section{introduction}

Quantum entanglement lies at the heart of quantum information
theory, with applications to quantum computing, teleportation, and
communication. It is widely accepted that understanding the
entanglement in a relativistic framework is not only of interest to
the quantum information, but also plays an important role in the
black hole physics \cite{Bombelli-Callen, Hawking-Terashima}. Thus,
much attention has been focus on the study of the quantum
information in a relativistic setting \cite{Peres,Boschi,
Bouwmeester, Alsing-Milburn, Alsing-McMahon-Milburn, Ge-Kim,
Schuller-Mann, Alsing-Mann, adesso, Ahn,Pan,Pan Qiyuan,Doukas}.
Recently, it was found that, due to the Unruh effect \cite{unruh},
the projective measurements made by the accelerated observer can
generate real particles detectable by the inertial observer both in
the cases of the scalar \cite{Han} and Dirac \cite{David} fields.
Furthermore, as shown in \cite{David}, the entanglement of the
produced state increase as the observer's acceleration increase.

As a further step along this line, the aim of this paper is to
investigate how the projective measurements made by the observer who
locates near the event horizon of the Schwarzschild black hole
affect the Dirac particle state detected by observer  who stays
stationary at the asymptotically flat region and how the Hawking
temperature \cite{Hawking-1} change the properties of the
entanglement. Our scheme can be set up by two observers, Alice and
Bob, together with their associated detectors. After the coincidence
of Alice and Bob at the same initial point in an asymptotically flat
region, Alice stays stationary at the flat place, while Bob falls
toward the Schwarzschild black hole and hovers at a fixed finite
nearest distance away from the horizon with uniform acceleration. As
a result of the Hawking effect, Bob will perceive a Fermi-Dirac
distribution of particles and antiparticles in the Kruskal vacuum.
Then we let Bob make a standard von Neumann projective measurement
(measuring the particle number)\cite{Han, David} and figure out what
Alice obtains. It is worth mentioning that if we let Bob always free
fall, the projective measurement made by him also have an affect on
Alice¡¯s state. However, this is in fact another issue because Bob
could not observes the existence of the event horizon.

The outline of the paper is as follows. In Sec. 2 we discuss the
features of quantum field theory in the Schwarzschild spacetime and
the Hawking effect for the Dirac particles. In Sec. 3 we analyze the
effects of the projective measurements on the generation of
entangled fermions in the Schwarzschild spacetime. We will summarize
and discuss our conclusions in the last section.

\section{Vacuum structure and Hawking Radiation for Dirac fields}

The metric for the Schwarzschild spacetime is given by
\begin{eqnarray}\label{matric}
ds^2=-\left(1-\frac{2M}{r}\right)
dt^2+\left(1-\frac{2M}{r}\right)^{-1} dr^2+r^2(d\theta^2
+sin^2\theta d\varphi^2),
\end{eqnarray}
where the parameter $M$ represents the mass of the black hole.
Throughout this paper we use $G=c=\hbar=\kappa_{B}=1$. Introducing a
tortoise coordinate
\begin{eqnarray}
r_{*}=r+2M\ln\frac{r-2M}{2M},
\end{eqnarray}
and defining the advanced time $v$ and retarded time $u$ as
\begin{eqnarray}
v=t+r_{*}\quad \quad\quad u=t-r_{*},
\end{eqnarray}
we obtain the generalized light-like Kruskal coordinates $\mathcal
{U}$ and $\mathcal {V}$ for the Schwarzschild black hole \cite{D-R}
\begin{eqnarray}
&&u=-4M\ln(-\frac{{\mathcal {U}}}{4M}),\quad v=4M\ln(\frac{{\mathcal
{V}}}{4M}),
\quad {\rm if ~~~r>r_{+}};\\
&&u=-4M\ln(\frac{{\mathcal {U}}}{4M}),\quad v=4M\ln(\frac{{\mathcal
{V}}}{4M}), \quad {\rm if~~~r<r_{+}};
\end{eqnarray}
where $\mathcal {U}$ and $\mathcal {V}$ are regular across the past
and future horizons of the extended spacetime.

For the Schwarzschild spacetime the Dirac equation \cite{Brill}
\begin{equation}\label{Di}
[\gamma^a e_a{}^\mu(\partial _\mu+\Gamma_\mu)]\Psi=0,
\end{equation}
can be written as
\begin{eqnarray}
\nonumber-\frac{\gamma_0}{\sqrt{1-\frac{2M}{r}}}\frac{\partial
\Psi}{\partial t}&&+\gamma_1\sqrt{1-\frac{2M}{r}}
\bigg[\frac{\partial }{\partial r}+\frac{1}{r}+\frac{M}{ 2r(r-2M)}
\bigg]\Psi\\&&+\frac{\gamma_2}{r}(\frac{\partial } {\partial
\theta}+\frac{1}{2}\cot\theta)\Psi+\frac{\gamma_3}{r
\sin\theta}\frac{\partial \Psi}{\partial \varphi}=0. \label{Di1}
\nonumber\\
\end{eqnarray}
If we re-scale $\Psi$ as
\begin{eqnarray}\Psi=(1-\frac{2M}{r})^{-\frac{1}{4}}\Phi,
\end{eqnarray} and use an
ansatz for the Dirac spinor
\begin{eqnarray}\label{C5an}
    \Phi=\left(
\begin{array}{c}
\frac{i \chi^{(\pm)}_{1}(r)}{r}\phi^{\pm}_{jm}(\theta, \varphi) \\
\frac{\chi^{(\pm)}_{2}(r)}{r}\phi^{\mp}_{jm}(\theta, \varphi)
\end{array}\right)e^{-i \omega t},
\end{eqnarray}
with spinor angular harmonics
\begin{eqnarray}\label{C54}
    \phi^{+}_{jm}=\left(
\begin{array}{c}
\sqrt{\frac{j+m}{2 j}}Y^{m-1/2}_l \\
\sqrt{\frac{j-m}{2 j}}Y^{m+1/2}_l
\end{array}\right), \ \ \ \ \ \ \ \ \ \ \ \  (for \ \
j=l+\frac{1}{2}),
\end{eqnarray}
\begin{eqnarray}
    \phi^{-}_{jm}=\left(
\begin{array}{c}
\sqrt{\frac{j+1-m}{2 j+2}}Y^{m-1/2}_l \\
-\sqrt{\frac{j+1+m}{2 j+2}}Y^{m+1/2}_l
\end{array}\right), \ \ \ \ \ \ (for \ \ j=l-\frac{1}{2}),
\end{eqnarray}
we find that the cases for $(+)$ and $(-)$ in the functions
$\chi^{(\pm)}_{1}$ and $\chi^{(\pm)}_{2}$ can be put together, and
then the decoupled equations can be expressed as
\begin{eqnarray}
    \frac{d^2 \chi_{1}}{d r_*^2}+(\omega^2-V_1)\chi_{1}&=&0, \label{even}\\
    \frac{d^2 \chi_{2}}{d r_*^2}+(\omega^2-V_2)\chi_{2}&=&0, \label{odd}
\end{eqnarray}
with
\begin{eqnarray}
V_1&=&\frac{\sqrt{1-\frac{2M}{r}}|K|}{r^2}\left(|K|\sqrt{1-\frac{2M}{r}}+\frac{r-3M}{r}\right),
\ \ \ \left( K=-j-\frac{1}{2},\ \ j=l-\frac{1}{2}
\right), \label{V1} \\
V_2&=&\frac{\sqrt{1-\frac{2M}{r}}|K|}{r^2}\left(|K|\sqrt{1-\frac{2M}{r}}-\frac{r-3M}{r}\right),
\ \ \ \left( K=j+\frac{1}{2},\ \ j=l+\frac{1}{2} \right). \label{V2}
\end{eqnarray}
Solving Eqs. (\ref{even}) and (\ref{odd}) near the event horizon, we
obtain $\chi_{1}=\chi_{2}=e^{\pm i\omega r_{*}}$. Hereafter we will
use the wavevector $\mathbf{k}$ labels the modes. Particles and
antiparticles will be classified with respect to the future-directed
timelike Killing vector in each region \cite{Alsing-Mann}. Thus, for
the outside and inside region of the event horizon, the positive
(fermions) frequency outgoing solutions are found to be \cite{D-R}
\begin{eqnarray}\label{outside mode}
\Psi^{I+}_{\mathbf{k}}=\mathcal {G}e^{-i\omega u},
\end{eqnarray}
\begin{eqnarray}\label{inside mode}
\Psi^{II+}_{\mathbf{k}}=\mathcal {G}e^{i\omega u},
\end{eqnarray}
where \begin{eqnarray} \mathcal{G}=\left[\begin{array}{c}
i(r^4-2Mr^3)^{-\frac{1}{4}}\phi^{\pm}_{jm}(\theta, \varphi) \\
(r^4-2Mr^3)^{-\frac{1}{4}}\phi^{\mp}_{jm}(\theta, \varphi)
\end{array}\right]\end{eqnarray} is a 4-component Dirac spinor.

Since the modes $\Psi^{I+}_{\mathbf{k}}$ and
$\Psi^{II+}_{\mathbf{k}}$ are analytic outside and inside the event
horizon respectively, they form a complete orthogonal family. Thus,
in terms of these bases the field  $\Psi_{out}$ can be expanded as
\begin{eqnarray}\label{First expand}
&&\Psi_{out}=\sum_{\sigma}\int
d\mathbf{k}[a^{\sigma}_{\mathbf{k}}\Psi^{\sigma+}_{\mathbf{k}}
+b^{\sigma\dag}_{\mathbf{k}}\Psi^{\sigma-}_{\mathbf{k}}],
\end{eqnarray}
where $\sigma=(I,II)$,  $a^{I}_{\mathbf{k}}$ and
$b^{I\dag}_{\mathbf{k}}$ are the fermion annihilation and
antifermion creation operators acting on the state of the exterior
region, and $a^{II}_{\mathbf{k}}$ and $b^{II\dag}_{\mathbf{k}}$ are
the fermion annihilation and antifermion creation operators acting
on the state of the interior region of the Schwarzschild black hole,
respectively.

On the other hand, by making an analytic continuation for Eqs.
(\ref{outside mode}) and (\ref{inside mode}), we find a complete
basis for positive energy modes  which analytic for all real
$\mathcal {U}$ and $\mathcal {V}$ according to the suggestion of
Domour-Ruffini \cite{D-R}
\begin{eqnarray}
\mathscr{F}^{I+}_{\mathbf{k}}=e^{2\pi
M\omega_\mathbf{k}}\Psi^{I+}_{\mathbf{k}} +e^{-2\pi
M\omega_\mathbf{k}}\Psi^{II-}_{-\mathbf{k}},
\end{eqnarray}
\begin{eqnarray}
\mathscr{F}^{II+}_{\mathbf{k}}=e^{-2\pi
M\omega_\mathbf{k}}\Psi^{I-}_{-\mathbf{k}} +e^{2\pi
M\omega_\mathbf{k}}\Psi^{II+}_{\mathbf{k}}.
\end{eqnarray}
Thus, we can also quantize the Dirac field in the Kruskal spacetime
as
\begin{eqnarray}\label{Second expand}
\Psi_{out}=\sum_{\sigma}\int d\mathbf{k}[2\cosh(4\pi
M\omega_\mathbf{k})]^{-1/2}[c^{\sigma}_{\mathbf{k}}
\mathscr{F}^{\sigma+}_{\mathbf{k}}+d^{\sigma\dag}_{\mathbf{k}}\mathscr{F}^{\sigma-}
_{\mathbf{k}} ],
\end{eqnarray}
where $c^{\sigma}_{\mathbf{k}}$ and $d^{\sigma\dag}_{\mathbf{k}}$
are the annihilation and creation operators acting on the Kruskal
vacuum.

Eqs. (\ref{First expand}) and (\ref{Second expand}) represent the
decomposition of the Dirac field in Schwarzschild  and Kruskal modes
respectively, so we can easily get the Bogoliubov transformations
\cite{Alsing-Mann}
\begin{eqnarray}
\left[
  \begin{array}{c}
    c^{\sigma}_{\mathbf{k}} \\
    d^{\sigma\dag}_{\mathbf{-k}} \\
  \end{array}
\right]=V\left[
  \begin{array}{c}
    a^{\sigma}_{\mathbf{k}} \\
    b^{\sigma\dag}_{\mathbf{-k}} \\
  \end{array}
\right]V^{-1},
\end{eqnarray}
where $V = \exp\left[r\,\left( a^{I\dagger}_\mathbf{k}
\,b^{II\dagger}_{-\mathbf{k}} \ +  a^{I}_\mathbf{k}
\,b^{II}_{-\mathbf{k}} \ \right)\right]$ is a two-mode Dirac
squeezing operator.

Then we assume that the Kruskal vacuum
$|0_{\mathbf{k}}\rangle^{+}_{\mathcal {K}}$ is related to the vacuum
of the black hole
$|0_{\mathbf{k}}\rangle^{+}_{I}|0_{-\mathbf{k}}\rangle ^{-}_{II}$ by
\begin{eqnarray}\label{two vacuum}
|0_\mathbf{k}\rangle^+_{\mathcal
{K}}=\digamma(a^{I}_{\mathbf{k}},a^{I\dag}_{\mathbf{k}},b^{II}
_{\mathbf{-k}},b^{II\dag}_{\mathbf{-k}})|0_{\mathbf{k}}\rangle^{+}_{I}|0_{-\mathbf{k}}\rangle
^{-}_{II}.
\end{eqnarray}
From $[a^{I}_{\mathbf{k}},a^{I\dag}_{\mathbf{k}}]=[b^{II}
_{\mathbf{-k}},b^{II\dag}_{\mathbf{-k}}]=1$ and
$c^{I}_{\mathbf{k}}|0_{\mathbf{k}}\rangle^{+}_{\mathcal {K}}=0$, we
get \cite{Pan Qiyuan,unruh,Ahn}
\begin{eqnarray}
\digamma\propto \exp\bigg(e^{-4\pi
M\omega_\mathbf{k}}a^{I\dag}_{\mathbf{k}}b^{II\dag}
_{\mathbf{-k}}\bigg).
\end{eqnarray}
After properly normalizing the state vector, the Kruskal particle
vacuum state for mode $\mathbf{k}$ is found to be
\begin{eqnarray}\label{Dirac-vacuum}
|0_{\mathbf{k}}\rangle^{+}_{\mathcal {K}}= (e^{-8\pi
M\omega_\mathbf{k}}+1)^{-\frac{1}{2}}\exp\bigg(e^{-4\pi
M\omega_\mathbf{k}}a^{I\dag}_{\mathbf{k}}b^{II\dag}
_{\mathbf{-k}}\bigg)|0_{\mathbf{k}}\rangle^{+}_{I}|0_{-\mathbf{k}}\rangle
^{-}_{II},
\end{eqnarray}
where $\{|n_{-\mathbf{k}}\rangle^{-}_{II}\}$ and
$\{|n_{\mathbf{k}}\rangle^{+}_{I}\}$ are the orthonormal bases for
the inside and outside region of the event horizon respectively, and
the $\{+,-\}$ superscript on the kets is used to indicate the
particle and antiparticle vacua.

A formal expression for the total Kruskal particle vacuum is
obtained by using Eq. (\ref{Dirac-vacuum}) for each mode
\begin{eqnarray}\label{all Dirac-vacuum}
|0\rangle^{+}_{\mathcal {K}}=\prod_{\mathbf{k}} (e^{-8\pi
M\omega_\mathbf{k}}+1)^{-\frac{1}{2}}\exp\bigg(e^{-4\pi
M\omega_\mathbf{k}}a^{I\dag}_{\mathbf{k}}b^{II\dag}
_{\mathbf{-k}}\bigg)|0_{\mathbf{k}}\rangle^{+}_{I}|0_{-\mathbf{k}}\rangle
^{-}_{II}.
\end{eqnarray}
Similarly, the total Kruskal antiparticle vacuum takes the form
\begin{eqnarray}\label{all antiDirac-vacuum}
|0\rangle^{-}_{\mathcal {K}}=\prod_{\mathbf{k}} (e^{-8\pi
M\omega_\mathbf{k}}+1)^{-\frac{1}{2}}\exp\bigg(-e^{-4\pi
M\omega_\mathbf{k}}b^{I\dag}_{\mathbf{k}}a^{II\dag}
_{\mathbf{-k}}\bigg)|0_{\mathbf{k}}\rangle^{-}_{I}|0_{-\mathbf{k}}\rangle
^{+}_{II}.
\end{eqnarray}
Then, the full Kruskal vacuum is $|0\rangle_{\mathcal {K}}=
|0\rangle^{+}_{\mathcal {K}}|0\rangle^{-}_{\mathcal {K}}$, which
corresponds to the absence of particles and antiparticles as
detected by the Kruskal observer.

When Bob travels through the Kruskal vacuum, his detector registers
the number of particles
\begin{eqnarray}\label{Hawking}
N^2_\omega=\frac{1}{e^{8\pi M\omega}+1}=\frac{1}{e^{\omega/T}+1},
\end{eqnarray}
which shows that the observer in the exterior of the black hole
detects a thermal Fermi-Dirac distribution of particles. Here we
have defined the Hawking temperature as $T=\frac{1}{8\pi M}$
\cite{Kerner-Mann,Jiang-Wu-Cai}.

\section{Generating entangled fermions by projective measurements}

We now study the relationships between projective measurements and
the generation of entangled particles in the background of the
Schwarzschild black hole.  The vacuum (\ref{all Dirac-vacuum}) could
be rewritten as
\begin{eqnarray}
\nonumber|0\rangle^{+}_{\mathcal
{K}}=&&(e^{-\omega_\mathbf{k}/T}+1)^{-\frac{1}{2}}\exp
\bigg(e^{-\frac{\omega_\mathbf{k}}{2T}}
a^{I\dag}_{\mathbf{k}}b^{II\dag}_{\mathbf{-k}}\bigg)\\
\nonumber&&\prod_{\mathbf{k'}\neq\mathbf{k}}
(e^{-\omega_\mathbf{k'}/T}+1)^{-\frac{1}{2}}\exp\bigg(e^{-\frac{\omega_\mathbf{k'}}{2T}}a^{I\dag}_{\mathbf{k'}}b^{II\dag}_{\mathbf{-k'}}\bigg)
|0_{\mathbf{k'}}\rangle^{+}_{I}|0_{-\mathbf{k'}}\rangle ^{-}_{II}
\end{eqnarray}
Now supposing that Bob performs a measurement on this state and
detects one particle in the mode $\mathbf{k}$, then the state will
be projected to the single particle state
\begin{eqnarray}
|\Phi_{+}(\mathbf{k})\rangle_B=\mathcal
{P}_{\mathbf{k}}\prod_{\mathbf{k'}}
(e^{-\omega_\mathbf{k'}/T}+1)^{-\frac{1}{2}}\exp\bigg(e^{-\frac{\omega_\mathbf{k'}}{2T}}
a^{I\dag}_{\mathbf{k'}}b^{II\dag}_{\mathbf{-k'}}\bigg)
|0_{\mathbf{k'}}\rangle^{+}_{I}|0_{-\mathbf{k'}}\rangle ^{-}_{II},
\end{eqnarray}
where $\mathcal {P}_{\mathbf{k}}$ is defined as
\begin{eqnarray}
\mathcal
{P}_{\mathbf{k}}=(e^{-\omega_\mathbf{k}/T}+1)^{\frac{1}{2}}a^{I\dag}_{\mathbf{k}}
b^{II\dag}_{\mathbf{-k}}
\exp\bigg(-e^{-\frac{\omega_\mathbf{k}}{2T}}a^{I\dag}
_{\mathbf{k}}b^{II\dag}_{\mathbf{-k}}\bigg).
\end{eqnarray}

Then we switch back to the Alice's frame to figure out what she
obtains \cite{Han,David}.  After some calculations for the
Bogoliubov transformation, we obtain
\begin{eqnarray}
\label{Kruskalpsik}
|\Phi_{+}(\mathbf{k})\rangle&=&V^{-1}|\Phi_{+}(\mathbf{k})\rangle_B\\
\nonumber&=&(e^{\omega_\mathbf{k}/T}+1)^{-\frac{1}{2}}
|\tilde{0}_\mathbf{k}\rangle^+ |\tilde{0}_\mathbf{-k}\rangle^-+
(e^{-\omega_\mathbf{k}/T}+1)^{-\frac{1}{2}}|\tilde{1}_\mathbf{k}\rangle^+
|\tilde{1}_\mathbf{-k}\rangle^-,
\end{eqnarray}
where $|\tilde{1}_\mathbf{k}\rangle^+$ and
$|\tilde{1}_\mathbf{-k}\rangle^-$ are single particle and
antiparticle excitation in Alice's detector. We see that from
Alice's perspective, the state is a superposition of the vacuum and
pair production of fermions an antifermions at frequency
$\omega_\mathbf{k}$. The physical mechanism of this process can be
given as follows. As a result of the Hawking effect, Bob will
perceive a Fermi-Dirac distribution of particles and antiparticles.
Since only a thermal bath in the accelerated frame corresponds to
vacuum obtains by Alice \cite{Han}, a measurement by Bob will
destroy the purely thermal nature of the field, then Alice can no
longer obtain a vacuum state.

In the case of a supermassive or an almost extreme black hole ($T
\rightarrow 0$), the state $|\Phi_{+}(\mathbf{k})\rangle$ approaches
to
\begin{eqnarray}
\lim_{T \rightarrow 0} |\Phi_+(\mathbf{k})\rangle =
|\tilde{1}_\mathbf{k}\rangle^+ |\tilde{1}_\mathbf{-k}\rangle^-.
\end{eqnarray}
It is interesting to note that the projective measurement made by
Bob also created particles detectable by Alice in the case of $T
\rightarrow 0$, but this particle state is unentangled in this case.

And in the limit $T \rightarrow \infty$, which corresponds to the
case of the black hole evaporates completely, the state
$|\Phi_{+}(\mathbf{k})\rangle$ approaches a maximally entangled Bell
state
\begin{eqnarray}
\lim_{T \rightarrow \infty} |\Phi_+(\mathbf{k})\rangle =
\frac{1}{\sqrt{2}}\bigg(|\tilde{0}_\mathbf{k}\rangle^+
|\tilde{0}_\mathbf{-k}\rangle^-+ |\tilde{1}_\mathbf{k}\rangle^+
|\tilde{1}_\mathbf{-k}\rangle^- \bigg).
\end{eqnarray}

It is well known that the degree of bipartite entanglement can be
quantified uniquely for pure states by the partial entropy of
entanglement, defined as the von Neumann entropy
\cite{Schuller-Mann,MA}
\begin{eqnarray}
E_{p}(|\psi\rangle)=S(\rho_A)=S(\rho_B)=-\sum_i \lambda_i
\log_2{\lambda_i}.
\end{eqnarray}
where $\rho_A$ (or $\rho_B$) is the reduced density matrix of
subsystem $A$ (or $B$), and $\lambda_i$ is the $i$th eigenvalues of
$\rho_A$ or $\rho_B$.

\begin{figure}[ht]
\includegraphics[scale=1.0]{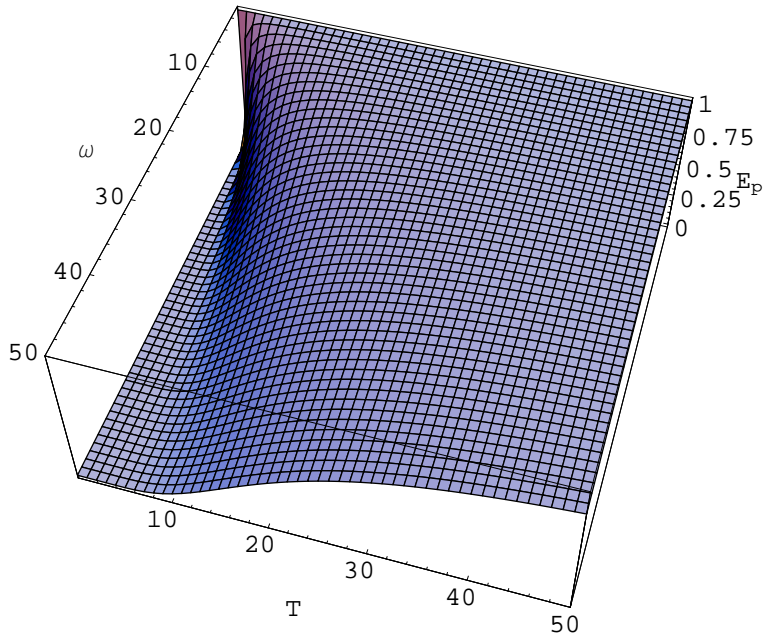}
\caption{\label{entropy}The degree of entanglement as a function of
the frequency of the detected particles $\omega_\mathbf{k}$ and the
Hawking temperature $T$. It demonstrates that only the lower
frequency particles are highly entangled. As the frequency
$\omega_\mathbf{k}\rightarrow\infty$, the degree of entanglement
decreases to zero.}
\end{figure}

The state (\ref{Kruskalpsik}) can be represented by the density
matrix
\begin{eqnarray}
&&\rho_\mathbf{k}=(e^{\omega_\mathbf{k}/T}+1)^{-1}|\tilde{0}\tilde{0}\rangle\langle\tilde{0}\tilde{0}|
+(e^{\omega_\mathbf{k}/T}+e^{-\omega_\mathbf{k}/T}+2)^{-\frac{1}{2}}|\tilde{0}\tilde{0}\rangle\langle\tilde{1}\tilde{1}|
\nonumber\\&&\qquad+(e^{\omega_\mathbf{k}/T}+e^{-\omega_\mathbf{k}/T}+2)^{-\frac{1}{2}}
|\tilde{1}\tilde{1}\rangle\langle\tilde{0}\tilde{0}|+
(e^{-\omega_\mathbf{k}/T}+1)^{-1}|\tilde{1}\tilde{1}\rangle\langle\tilde{1}\tilde{1}|,
\end{eqnarray}
where $|\tilde{n}\tilde{m}\rangle=|\tilde{n}_\mathbf{k}\rangle^+
\otimes|\tilde{m}_\mathbf{-k}\rangle^-$. In order to find the
partial entropy  of the state, we trace out one subsystem of the
density matrix and obtain
\begin{eqnarray}
\rho^{+}_\mathbf{k}=Tr_{-}(\rho_\mathbf{k})=
(e^{\omega_\mathbf{k}/T}+1)^{-1}|\tilde{0}\rangle\langle\tilde{0}|+
(e^{-\omega_\mathbf{k}/T}+1)^{-1}|\tilde{1}\rangle\langle\tilde{1}|,
\end{eqnarray}
where $|\tilde{n}\rangle=|\tilde{n}_\mathbf{k}\rangle^+$. From which
the partial entropy of entanglement is found to be
\begin{eqnarray}
E_{p}[|\Phi_+(\mathbf{k})\rangle]= \log_2(e^{\omega_\mathbf{k}/T}+1)
+ (e^{-\omega_\mathbf{k}/T}+1)^{-1}
\log_2(e^{-\omega_\mathbf{k}/T}).
\end{eqnarray}
The behaviors of the partial entropy of entanglement as a function
of the frequency of the detected particles and the Hawking
temperature $T$ are illustrated Fig.\ref{entropy}. We find that, in
the case of a supermassive or an almost extreme black hole
($T\rightarrow0$), the degree of entanglement is zero, which
verifies that the state is an uncorrelated pure state.  The state
always entangled for nonzero Hawking temperature, and higher Hawking
temperature producing more entanglement. As $T\rightarrow\infty$,
corresponding to the case of the black hole evaporates completely,
the degree of entanglement is $E_{p}[|\Phi_+(\mathbf{k})\rangle]=1$,
which just verifies that the state in this case is the maximally
entangled Bell state.

If Bob detects an antiparticle in mode $\mathbf{k}$, the resulting
state can be simplifies to
\begin{eqnarray}
\label{anKruskalpsik}
|\Phi_{-}(\mathbf{k})\rangle=\bigg[(e^{\omega_\mathbf{k}/T}+1)^{-\frac{1}{2}}-
(e^{-\omega_\mathbf{k}/T}+1)^{-\frac{1}{2}}d^{I\dag}_{\mathbf{k}}c^{II\dag}_
{\mathbf{-k}}\bigg]|\tilde{0}_\mathbf{k}\rangle^-
|\tilde{0}_\mathbf{-k}\rangle^+,
\end{eqnarray}
which is entangled in the occupation number of the antiparticle mode
$\mathbf{k}$ and the particle mode $\mathbf{-k}$. This state also
approaches a new Bell state
\begin{eqnarray}
\lim_{T \rightarrow
\infty}|\Phi_{-}(\mathbf{k})\rangle=\frac{1}{\sqrt{2}}\bigg(|\tilde{0}_\mathbf{k}\rangle^-
|\tilde{0}_\mathbf{-k}\rangle^+ + |\tilde{1}_\mathbf{k}\rangle^-
|\tilde{1}_\mathbf{-k}\rangle^+\bigg),
\end{eqnarray}
when the black hole evaporates completely. And it approaches the
uncorrelated pure state
\begin{eqnarray}
\lim_{T \rightarrow
0}|\Phi_{-}(\mathbf{k})\rangle=|\tilde{1}_\mathbf{k}\rangle^-
|\tilde{1}_\mathbf{-k}\rangle^+
\end{eqnarray}
in the case of supermassive or an almost extreme black hole.

\section{summary}

Building on the well-known Hawking effect, we have discussed the
effects of projective measurements on the generation of entangled
particles between two Dirac modes. It is shown that the projective
measurements made by Bob who falls toward a Schwarzschild black hole
will create entangled particles detected by Alice who stays
stationary at the asymptotically flat region. The physical mechanism
of this process is that once Bob makes a measurement, he destroys
the purely thermal nature of the field, then Alice no longer obtains
vacuum. We have demonstrated that only the lower frequency particles
are highly entangled. As the frequency
$\omega_\mathbf{k}\rightarrow\infty$, the degree of entanglement
decreases to zero, that is to say, the state is an unentangled pure
state. It is interesting to note that, when the Hawking temperature
is zero, i.e., the case of supermassive or an almost extreme black
hole, the projective measurement also created particles detectable
by Alice, but the state is unentangled in this case. For nonzero
Hawking temperature, the state is always entangled and its degree of
entanglement increases as the Hawking temperature increases. In the
limit of infinite Hawking temperature, i.e., the black hole
evaporates completely, the degree of entanglement is exactly one,
which just indicates that a maximally entangled Bell state is
produced.

\begin{acknowledgments}

This work was supported by the National Natural Science Foundation
of China under Grant No 10875040; the key project of the National
Natural Science Foundation of China under Grant No 10935013; the
National Basic Research of China under Grant No. 2010CB833004, the
Hunan Provincial Natural Science Foundation of China under Grant No.
08JJ3010, PCSIRT, No. IRT0964, and Construct Program of the National
Key Discipline.

\end{acknowledgments}

\end{document}